\documentstyle[12pt]{article}

\newcommand {\nn}    {\nonumber}
\newcommand {\vs}[1]  { \vspace*{#1 cm} }
\newcounter{eq}
\newcounter{sc}

\newcommand {\PL}   {Phys.Lett.}
\newcommand {\PR}   {Phys.Rev.}
\newcommand {\PRL}   {Phys.Rev.Lett.}

\newcommand {\PREP}  {Phys.Rep.}
\def\overleftrightarrow#1{\vbox{\ialign{##\crcr
 $\leftrightarrow$\crcr\noalign{\kern-1pt\nointerlineskip}
 $\hfil\displaystyle{#1}\hfil$\crcr}}}

\newlength{\minitwocolumn}
\begin{document}

%%%%%%%%%%%%%%%%%%%%%%%%%%%%%%%%%%%%%%%%%%%%%%%%%%%%%%%%%%%%%%%%%%
%%%%%%%%%%%%%%%%%%%%%%%% Title %%%%%%%%%%%%%%%%%%%%%%%%%%%%%%%%%%%
%%%%%%%%%%%%%%%%%%%%%%%%%%%%%%%%%%%%%%%%%%%%%%%%%%%%%%%%%%%%%%%%%%

\begin{flushright}
EDO-EP-33\\
August, 2000\\
\end{flushright}
\vspace{30pt}

%\magnification=\magstep1
\pagestyle{empty}
\baselineskip15pt
%\font\cmssB=cmss17
%\font\cmssS=cmss10

\begin{center}
{\large\bf Localization of Gravitino on a Brane

 \vskip 1mm
}

\vspace{20mm}

Ichiro Oda
          \footnote{
          E-mail address:\ ioda@edogawa-u.ac.jp
                  }
\\
\vspace{10mm}
          Edogawa University,
          474 Komaki, Nagareyama City, Chiba 270-0198, JAPAN \\

\end{center}

%\maketitle

\vspace{15mm}
\begin{abstract}
We show how the spin 3/2 gravitino field can be localized on a brane 
in a general framework of supergravity theory.
Provided that a scalar field coupled to the Rarita-Schwinger field
develops an vacuum expectation value (VEV) whose phase depends on 
the 'radial' coordinate in extra internal space, the gravitino
is localized on a brane with the exponentially decreasing warp factor
by selecting an appropriate value of the VEV.

\vspace{15mm}

\end{abstract}

\newpage
\pagestyle{plain}
\pagenumbering{arabic}
%\setcounter{page}{1}

%%%%%%%%%%%%%%%%%%%%%%%%%%%%%%%%%%%%%%%%%%%%%%%%%%%%%%%%%%%%%%%%%%
%%%%%%%%%%%%%%%%%%%%%%%% Article %%%%%%%%%%%%%%%%%%%%%%%%%%%%%%%%%
%%%%%%%%%%%%%%%%%%%%%%%%%%%%%%%%%%%%%%%%%%%%%%%%%%%%%%%%%%%%%%%%%%

\rm
%%%%%%%%%%%%%%%%%%%%%%%%%%%%%%%%%%%%%%%%%%%%%%%%%%%%%%%%%%%%%%%%%%%%%
%%%%%%%%%%%%%%%%%%%%%%%%%%%%%%   SEC  1    %%%%%%%%%%%%%%%%%%%%%%%%%%
%%%%%%%%%%%%%%%%%%%%%%%%%%%%%%%%%%%%%%%%%%%%%%%%%%%%%%%%%%%%%%%%%%%%%
\section{Introduction}

The fact that the three spatial dimensions we observe at present are 
expanding, and once were highly curved provides us a logical possibility
that there are other dimensions which not only remain small and highly
curved but also extend widely and are almost flat. The latter case might
appear to conflict with the Newton's law of gravity since the gravitational
interaction sees a whole structure of space-time so that the Newton's law
depends on crucially the dimensionality of space. In recent years, 
however, it has been shown that we can construct a gravity-localized model
by putting a 3-brane in $AdS_5$ \cite{Randall1, Randall2}. 
Afterwards, the model with a single or two branes \cite{Randall1, Randall2}
was generalized to a model with many domain walls on topology 
$S^1$ \cite{Oda1, Oda2}.

One interesting possibility in the gravity localized models is that various
spin fields as well as the spin 2 graviton might be confined to
a 3-brane. Indeed, this possibility has been recently examined in the 
Randall-Sundrum model \cite{Randall1, Randall2} by
a field theoretic approach in detail \cite{Rizzo, Pomarol, Grossman, Bajc}
where we have seen that spin 0 and 2 fields are localized on a brane while
spin 1/2, 1 and 3/2 are not. Thus, for a complete localization of all the 
fields on the brane it is necessary to invoke the other interactions in 
addition to the gravitational one in a bulk. For instance, in Ref.
\cite{Bajc}, 
the Dvali-Shifman mechanism \cite{Dvali} for the spin 1 vector field and 
the Jackiw-Rebbi mechanism \cite{Jackiw} for the fermionic fields were
proposed. 

There has been also a lot of activities of constructing
some higher dimensional generalizations of the Randall-Sundrum model in five 
dimensions \cite{Chodos, Cohen, Gregory, Chacko, Csaki, Chen,
Vilenkin, Gherghetta1, Chaichian, Chodos2, Oda3, Gherghetta2, Oda4, 
Hayakawa}. About the localization of various fields in those generalizations, 
it was shown that fermions are not localized on a brane either as in the
Randall-Sundrum
model \cite{Oda3}.     

More recently, a mechanism for localizing fermions on a brane in the context
of higher dimensional gravity localized models has been presented
\cite{Randjbar}. This mechanism makes use of a Yukawa coupling to a scalar
field 
and makes it possible to have chiral fermions as well as the localization
of spin
1/2 fermions on a brane.

The aim of the present article is to extend this mechanism for the spin 1/2
fermion to the case of the spin 3/2 gravitino field in a general framework of
supergravity theory. We will see that our mechanism provides not only
the localization but also the chirality of the gravitino as in the spin 1/2
fermions. It is worthwhile to emphasize that our presentation is quite 
general in the sense that we treat the most general coupling to supergravity
of chiral superfields describing complex scalar fields and Weyl spinor fields.

Thus, a complete localization of all spin fields on a brane is now available
from the viewpoint of the local field theory. Namely, in the gravity localized
models in an arbitrary space-time dimension except six dimensions, spin 0
and 2 fields are localized on a brane with the exponentially decreasing
warp factor by only the gravitational interaction, spin 1 vector field
is by the Dvali-Shifman mechanism, and spin 1/2 and 3/2 fermionic fields
are so through a coupling to a scalar field. On the other hand, in a
string-like
defect model with codimension 2 in six dimensions \cite{Gregory,
Gherghetta1}, 
the spin 1 vector field is
also localized on a brane by the gravitational interaction so we do not
have to appeal to the Dvali-Shifman mechanism \cite{Oda3}.

%%%%%%%%%%%%%%%%%%%%%%%%%%%%%%%%%%%%%%%%%%%%%%%%%%%%%%%%%%%%%%%%%%%%%
%%%%%%%%%%%%%%%%%%%%%%%%%%%%%%   SEC  2    %%%%%%%%%%%%%%%%%%%%%%%%%%
%%%%%%%%%%%%%%%%%%%%%%%%%%%%%%%%%%%%%%%%%%%%%%%%%%%%%%%%%%%%%%%%%%%%%
\section{The model}

Let us start by considering the general supergravity action
coupled to chiral supermultiplets in $D$ dimensions \cite{Nilles}
(Only the relevant part for the present consideration in the total 
action is explicitly written down.):
%**   1 %%%%%%%%%%%%%%%%%%%%%%%%%%%%%%%%%%%%%%%%%%%%%%%%%%%%%%%%%
\begin{eqnarray}
S = \int d^D x  
\sqrt{-g} [ \frac{1}{2 \kappa_D^2} R +
\bar{\Psi}_M i \Gamma^{[M} \Gamma^N \Gamma^{R]} 
\left\{ D_N \Psi_R + i \lambda \Gamma^r ( G^i D_N \phi_i
- G_i D_N \phi^{*i} ) \Psi_R \right\}  \nn\\
- G^i_j g^{MN} D_M \phi_i D_N \phi^{*j} 
+ \cdots ],
\label{1}
\end{eqnarray}
%%%%%%%%%%%%%%%%%%%%%%%%%%%%%%%%%%%%%%%%%%%%%%%%%%%%%%%%%%%%%%%%%%%
where $\kappa_D$ and $\lambda$ respectively denote the $D$-dimensional 
gravitational constant and a coupling constant, $M, N, R, \cdots$ 
do $D$-dimensional indices,
and the square bracket on the curved gamma matrices $\Gamma^M$
denotes the anti-symmetrization with weight 1. 
The Rarita-Schwinger gravitino field in the gravity multiplet and 
complex scalar fields in the chiral superfields 
are respectively described as ${\Psi}_M$ and $\phi_i$. 
(Here the indices $i, j$ denote the number of fields.)  
Moreover, the function $G$, which is usually called to
the Kahler potential, is defined as
%**   2 %%%%%%%%%%%%%%%%%%%%%%%%%%%%%%%%%%%%%%%%%%%%%%%%%%%%%%%%%
\begin{eqnarray}
G(\phi^*, \phi) = -3 \log (-\frac{K(\phi^*, \phi)}{3})
+ \log|W(\phi)|^2,
\label{2}
\end{eqnarray}
%%%%%%%%%%%%%%%%%%%%%%%%%%%%%%%%%%%%%%%%%%%%%%%%%%%%%%%%%%%%%%%%%%%
with a general function $K$ and the superpotential $W$.
Various derivatives of the Kahler potential are described as
%**   3 %%%%%%%%%%%%%%%%%%%%%%%%%%%%%%%%%%%%%%%%%%%%%%%%%%%%%%%%%
\begin{eqnarray}
G^i = \frac{\partial G}{\partial \phi_i}, \  G_i = \frac{\partial G}
{\partial \phi^{*i}}, \ G^i_j = \frac{\partial^2 G}
{\partial \phi_i \partial \phi^{*j}}.
\label{3}
\end{eqnarray}
%%%%%%%%%%%%%%%%%%%%%%%%%%%%%%%%%%%%%%%%%%%%%%%%%%%%%%%%%%%%%%%%%%%
And the covariant derivative $D_M \Psi_N$ is defined as
%**   4 %%%%%%%%%%%%%%%%%%%%%%%%%%%%%%%%%%%%%%%%%%%%%%%%%%%%%%%%%
\begin{eqnarray}
D_M \Psi_N = \partial_M \Psi_N - \Gamma^R_{MN} \Psi_R 
+ \frac{1}{4} \omega_M^{AB} \Gamma_{AB} \Psi_N,
\label{4}
\end{eqnarray}
%%%%%%%%%%%%%%%%%%%%%%%%%%%%%%%%%%%%%%%%%%%%%%%%%%%%%%%%%%%%%%%%%%%
where the spin connection $\omega_M^{AB}$ and the affine connection 
$\Gamma^R_{MN}$ are defined as usual, and the indices $A, B$ denote
the $D$-dimensional local Lorentz indices. As a final remark,
throughout this article we follow the standard conventions and 
notations of the textbook of Misner, Thorne and Wheeler \cite{Misner}. 

In this note, without loss of generality, we shall choose the form of $K$
which leads to a minimal kinetic term and set the number of complex 
scalar fields to one. Then, 
%**   5 %%%%%%%%%%%%%%%%%%%%%%%%%%%%%%%%%%%%%%%%%%%%%%%%%%%%%%%%%
\begin{eqnarray}
K &=& -3 \exp(-\frac{\phi \phi^*}{3}), \nn\\
G &=& \phi \phi^* + \log|W(\phi)|^2, \nn\\ 
G^i_j &=& 1.
\label{5}
\end{eqnarray}
%%%%%%%%%%%%%%%%%%%%%%%%%%%%%%%%%%%%%%%%%%%%%%%%%%%%%%%%%%%%%%%%%%%
{}From the action (\ref{1}), the equations of motion for the 
Rarita-Schwinger gravitino field are derived as
%**   6 %%%%%%%%%%%%%%%%%%%%%%%%%%%%%%%%%%%%%%%%%%%%%%%%%%%%%%%%%
\begin{eqnarray}
\Gamma^{[M} \Gamma^N \Gamma^{R]} \left(D_N + i \lambda \Gamma^r
\phi^* \overleftrightarrow{\partial_N} \phi \right) \Psi_R = 0,
\label{6}
\end{eqnarray}
%%%%%%%%%%%%%%%%%%%%%%%%%%%%%%%%%%%%%%%%%%%%%%%%%%%%%%%%%%%%%%%%%%%
with the definition of $\phi^* \overleftrightarrow{\partial_N} \phi
= \phi^* \partial_N \phi - \partial_N \phi^* \phi$.

We shall adopt the following cylindrically symmetric metric ansatz:
%**   7 %%%%%%%%%%%%%%%%%%%%%%%%%%%%%%%%%%%%%%%%%%%%%%%%%%%%%%%%%
\begin{eqnarray}
ds^2 &=& g_{MN} dx^M dx^N  \nn\\
&=& e^{-A(r)} \hat{g}_{\mu\nu}(x^\lambda) dx^\mu dx^\nu + dr^2 
+ e^{-B(r)} \hat{g}_{mn}(y) dy^m dy^n,
\label{7}
\end{eqnarray}
%%%%%%%%%%%%%%%%%%%%%%%%%%%%%%%%%%%%%%%%%%%%%%%%%%%%%%%%%%%%%%%%%%%
where as mentioned above $M, N, ...$ denote $D$-dimensional space-time 
indices, $\mu, \nu, ...$ do $D_1$-dimensional brane ones, 
and $m, n, ...$ do $D_2$-dimensional extra spatial ones, 
so the equality $D = D_1 + D_2 +1$ holds. 
The local Lorentz indices corresponding to curved indices
$M = (\mu, r, m)$ are described as $A = (a, r, \underline{a})$.
For simplicity, from now on we limit ourselves to the flat brane geometry 
$\hat{g}_{\mu\nu} = \eta_{\mu\nu}$.
With this ansatz, after a straightforward calculation, the components 
of the covariant derivative are easily calculated:
%**   8 %%%%%%%%%%%%%%%%%%%%%%%%%%%%%%%%%%%%%%%%%%%%%%%%%%%%%%%%%
\begin{eqnarray}
D_\mu \Psi_\nu &=& (\partial_\mu + \frac{1}{4} A' \Gamma_r
\Gamma_\mu) \Psi_\nu - \frac{1}{2} A' e^{-A} \eta_{\mu\nu} \Psi_r, \nn\\
D_\mu \Psi_r &=& (\partial_\mu + \frac{1}{4} A' \Gamma_r
\Gamma_\mu) \Psi_r + \frac{1}{2} A' \Psi_\mu, \nn\\
D_\mu \Psi_m &=& (\partial_\mu + \frac{1}{4} A' \Gamma_r
\Gamma_\mu) \Psi_m, \nn\\
D_r \Psi_\mu &=& \partial_r \Psi_\mu + \frac{1}{2} A' \Psi_\mu, \nn\\
D_r \Psi_r &=& \partial_r \Psi_r, \nn\\
D_r \Psi_m &=& \partial_r \Psi_m + \frac{1}{2} B' \Psi_m, \nn\\
D_m \Psi_\mu &=& (\partial_m + \frac{1}{4} B' \Gamma_r \Gamma_m 
+ \omega_m ) \Psi_\mu, \nn\\
D_m \Psi_r &=& (\partial_m + \frac{1}{4} B' \Gamma_r \Gamma_m 
+ \omega_m ) \Psi_r + \frac{1}{2} B' \Psi_m, \nn\\
D_m \Psi_n &=& (\hat{D}_m + \frac{1}{4} B' \Gamma_r \Gamma_m) \Psi_n
- \frac{1}{2} B' g_{mn} \Psi_r, 
\label{8}
\end{eqnarray}
%%%%%%%%%%%%%%%%%%%%%%%%%%%%%%%%%%%%%%%%%%%%%%%%%%%%%%%%%%%%%%%%%%%
where the prime denotes the differentiation with respect to $r$,
and $\hat{D}_m \Psi_n \equiv (\partial_m + \omega_m) \Psi_n + 
\hat{\Gamma}^p_{mn} \Psi_p$ with the spin connection $\omega_m$ and
the affine one $\hat{\Gamma}^p_{mn}$ made in terms of $\hat{g}_{mn}$ .

%%%%%%%%%%%%%%%%%%%%%%%%%%%%%%%%%%%%%%%%%%%%%%%%%%%%%%%%%%%%%%%%%%%%%
%%%%%%%%%%%%%%%%%%%%%%%%%%%%%%   SEC  3    %%%%%%%%%%%%%%%%%%%%%%%%%%
%%%%%%%%%%%%%%%%%%%%%%%%%%%%%%%%%%%%%%%%%%%%%%%%%%%%%%%%%%%%%%%%%%%%%
\section{Localization and chirality}

{}For spontaneous supersymmetry breaking to occur, at least
one of the scalar fields must have a vacuum expectation value (VEV)
which is not invariant under local supersymmetry transformations.
In this note, we assume that a single scalar field in our model
would develop such a VEV with the $r$-dependent phase factor
%**   9 %%%%%%%%%%%%%%%%%%%%%%%%%%%%%%%%%%%%%%%%%%%%%%%%%%%%%%%%%
\begin{eqnarray}
\phi = |v| e^{i |r|},
\label{9}
\end{eqnarray}
%%%%%%%%%%%%%%%%%%%%%%%%%%%%%%%%%%%%%%%%%%%%%%%%%%%%%%%%%%%%%%%%%%%
where $v$ is a constant. Then we have an equation
%**   10 %%%%%%%%%%%%%%%%%%%%%%%%%%%%%%%%%%%%%%%%%%%%%%%%%%%%%%%%%
\begin{eqnarray}
-i \lambda \phi^* \overleftrightarrow{\partial_r} \phi
= 2 \lambda |v|^2 \varepsilon(r).
\label{10}
\end{eqnarray}
%%%%%%%%%%%%%%%%%%%%%%%%%%%%%%%%%%%%%%%%%%%%%%%%%%%%%%%%%%%%%%%%%%%

With the assumption $\Psi_r = \Psi_m = 0$ \footnote{This assumption
might be legitimate since these Kaluza-Klein spin 1/2 fields would
have an infinite energy so they should be unphysical in a noncompact
extra space as shown in the 'gauge-scalar' coming from the KK reduction 
of the vector field \cite{Oda4}.}  
and using Eqs. (\ref{8}) and (\ref{10}), 
Eq. (\ref{6}) reduces to
%**   11 %%%%%%%%%%%%%%%%%%%%%%%%%%%%%%%%%%%%%%%%%%%%%%%%%%%%%%%%%
\begin{eqnarray}
\left[ \Gamma^r \left(\partial_r -  \frac{D_1-2}{4} A' - 
\frac{D_2}{4} B' \right) + 2 \lambda |v|^2 \varepsilon(r)
+ \Gamma^m (\partial_m + \omega_m) \right] \Psi_\mu = 0.
\label{11}
\end{eqnarray}
%%%%%%%%%%%%%%%%%%%%%%%%%%%%%%%%%%%%%%%%%%%%%%%%%%%%%%%%%%%%%%%%%%%
Here we have used $D_1$-dimensional equations $\gamma^\mu \Psi_\mu
= \partial^\mu \Psi_\mu = \gamma^{[\mu} \gamma^\nu \gamma^{\rho]} 
\partial_\nu \Psi_\rho = 0$ where we have defined $\gamma^\mu$ by
$\Gamma^\mu = e^{\frac{1}{2}A} \gamma^\mu$. Note that 
$\{\gamma^\mu, \gamma^\nu\} = 2 \eta^{\mu\nu}$.
Now let us look for a solution with
the form of $\Psi_\mu(x^M) = \psi_\mu(x^\mu) u(r) \chi(y^m)$ where
we impose the following chirality condition and field equation:
%**   12 %%%%%%%%%%%%%%%%%%%%%%%%%%%%%%%%%%%%%%%%%%%%%%%%%%%%%%%%%
\begin{eqnarray}
\Gamma^r \psi_\mu(x) = + \psi_\mu(x), \ \Gamma^m (\partial_m + \omega_m) 
\chi(y) = 0.
\label{12}
\end{eqnarray}
%%%%%%%%%%%%%%%%%%%%%%%%%%%%%%%%%%%%%%%%%%%%%%%%%%%%%%%%%%%%%%%%%%%
Then, Eq. (\ref{11}) is of the form
%**   13 %%%%%%%%%%%%%%%%%%%%%%%%%%%%%%%%%%%%%%%%%%%%%%%%%%%%%%%%%
\begin{eqnarray}
\left(\partial_r -  \frac{D_1-2}{4} A' - \frac{D_2}{4} B' 
+ 2 \lambda |v|^2 \varepsilon(r) \right) u(r) = 0,
\label{13}
\end{eqnarray}
%%%%%%%%%%%%%%%%%%%%%%%%%%%%%%%%%%%%%%%%%%%%%%%%%%%%%%%%%%%%%%%%%%%
{}from which, $u(r)$ is solved to be
%**   14 %%%%%%%%%%%%%%%%%%%%%%%%%%%%%%%%%%%%%%%%%%%%%%%%%%%%%%%%%
\begin{eqnarray}
u(r) = c_1 e^{\frac{D_1-2}{4} A + \frac{D_2}{4} B 
- 2 \lambda |v|^2 \varepsilon(r) r},
\label{14}
\end{eqnarray}
%%%%%%%%%%%%%%%%%%%%%%%%%%%%%%%%%%%%%%%%%%%%%%%%%%%%%%%%%%%%%%%%%%%
with $c_1$ being an integration constant.

We are now ready to show that the solution (\ref{14}) is normalizable
and consequently we have the localized gravitino on a brane if we 
take a suitable VEV. To this end, let us focus on the kinetic term
of the Rarita-Schwinger field in the action (\ref{1}) and substitute
the solution (\ref{14}) into it as follows:
%**   15 %%%%%%%%%%%%%%%%%%%%%%%%%%%%%%%%%%%%%%%%%%%%%%%%%%%%%%%%%
\begin{eqnarray}
S_{3/2} &\equiv& \int d^D x \sqrt{-g} \bar{\Psi}_M i \Gamma^{[M} 
\Gamma^N \Gamma^{R]} D_N \Psi_R \nn\\
&=& \int d^{D_2}y \sqrt{\hat{g}_y} \chi^{\dagger}(y) \chi(y)
\int dr e^{-\frac{1}{2}(D_1 A + D_2 B) + \frac{3}{2}A} u(r)^2 
\int d^{D_1}x \bar{\psi}_\mu i \gamma^{[\mu} 
\gamma^\nu \gamma^{\rho]} \partial_\nu \psi_\rho + \cdots.
\label{15}
\end{eqnarray}
%%%%%%%%%%%%%%%%%%%%%%%%%%%%%%%%%%%%%%%%%%%%%%%%%%%%%%%%%%%%%%%%%%%
Provided that the first integral over the extra internal space is finite,
the condition that $S_{3/2}$ has a normalizable kinetic term requires 
that the second integral should be finite. If we denote this integral
as $I$, we have an equation
%**   69 %%%%%%%%%%%%%%%%%%%%%%%%%%%%%%%%%%%%%%%%%%%%%%%%%%%%%%%%%
\begin{eqnarray}
I &\equiv& \int dr e^{-\frac{1}{2}(D_1 A + D_2 B) + \frac{3}{2}A} u(r)^2  
\nn\\
&=& 2 c_1^2 \int_0^{\infty} dr e^{\frac{1}{2} A - 4 \lambda |v|^2 r} \nn\\
&=& 2 c_1^2 \int_0^{\infty} dr e^{(\frac{1}{2} c - 4 \lambda |v|^2) r},
\label{69}
\end{eqnarray}
%%%%%%%%%%%%%%%%%%%%%%%%%%%%%%%%%%%%%%%%%%%%%%%%%%%%%%%%%%%%%%%%%%%
where we have considered a warp geometry $A = cr (c>0)$ 
in the last equation. Then it turns out that $I$ is finite
if a VEV $|v|$ has a large value such that $\frac{1}{2} c - 4 \lambda |v|^2
< 0$. Note that in the absence of a VEV of a scalar field
the solution is not normalizable so that the gravitino is 
localized not on a brane with the exponentially decreasing warp factor
but on a brane with the exponentially rising warp factor \cite{Bajc, Oda3}. 
In the above, we have imposed the chiral 
condition on the brane gravitino $\psi_\mu$, and then the number
of chiral families would depend on the topology of the extra
internal space \cite{Randjbar}.

%%%%%%%%%%%%%%%%%%%%%%%%%%%%%%%%%%%%%%%%%%%%%%%%%%%%%%%%%%%%%%%%%%%%%
%%%%%%%%%%%%%%%%%%%%%%%%%%%%%%   SEC  4    %%%%%%%%%%%%%%%%%%%%%%%%%%
%%%%%%%%%%%%%%%%%%%%%%%%%%%%%%%%%%%%%%%%%%%%%%%%%%%%%%%%%%%%%%%%%%%%%
\section{Discussions}

In this paper, within a general framework of supergravity,
we have investigated the possibility of localizing
the spin 3/2 gravitino field on a brane with the exponentially
decreasing warp factor, which also localizes the graviton.
The mechanism that we have found is quite similar to that 
found recently in case of the spin 1/2 fermion by Randjbar-Daemi
and Shaposhnikov \cite{Randjbar}. But one important difference
is that we have made use of spontaneous supersymmetry breaking
in the bulk in a general supergravity coupled to chiral superfields. 
The brane gravitino remains to be
massless and chiral after bulk-SUSY breaking, so it appears that
supersymmetry on a brane is not broken. Of course, to confirm
this statement we need further study in future.

By extensive study done so far about the localization of various spin fields
on a brane, we could conclude that all the Standard Model particles
(which include the graviton and the gravitino) can be confined
to a 3-brane by means of a local field theory, so it seems to be
timely in near future to construct a more realistic, phenomenological model 
based on the gravity localized models.

\vs 1
%%%%%%%%%%%%%%%%%%%%%%%%%%%%%%%%%%%%%%%%%%%%%%%%%%%%%%%%%%%%%%%%%%
%%%%%%%%%%%%%%%%%%%%%%%% reference %%%%%%%%%%%%%%%%%%%%%%%%%%%%%%%
%%%%%%%%%%%%%%%%%%%%%%%%%%%%%%%%%%%%%%%%%%%%%%%%%%%%%%%%%%%%%%%%%%

\end{document}